\documentclass[twocolumn,prl,aps,showpacs]{revtex4}
\usepackage{graphicx}
\usepackage{amsmath}
\usepackage{color}

\renewcommand{\vec}[1]{\boldsymbol{#1}}

\begin{document}

\title{Critical Flow and Dissipation in a Quasi-One-Dimensional Superfluid}

\date{\today}
\author{P-F Duc$^{1}$, M.Savard$^{1}$ , M. Petrescu$^{1}$, B. Rosenow$^{2}$, A. Del Maestro$^{3}$,  and G. Gervais$^{1,4}$}
\affiliation{$^{1}$Department of Physics, McGill University, Montreal, QC, H3A 2T8 Canada}
\affiliation{$^{2}$Institut f\"ur Theoretische Physik, Universit\"at Leipzig, D-04103, Leipzig, Germany}
\affiliation{$^{3}$Department of Physics, University of Vermont, Burlington, VT USA}
\affiliation{$^{4}$Canadian Institute for Advanced Research, Toronto, Canada}
\affiliation{$^{*}$corresponding author: gervais@physics.mcgill.ca}

\begin{abstract}
\vspace*{0.2cm}
In one of the most celebrated examples of the theory of universal critical phenomena, the phase transition to the superfluid state of $^{4}$He belongs to the same three dimensional $\mathrm{O}(2)$ universality class as the onset of ferromagnetism in a lattice of classical spins with $XY$ symmetry. Below the transition, the superfluid density $\rho_s$ and superfluid velocity $v_s$ increase as power laws of temperature described by a universal critical exponent constrained to be equal by scale invariance. As the dimensionality is reduced towards one dimension (1D), it is expected that enhanced thermal and quantum fluctuations preclude long-range order, thereby inhibiting superfluidity. We have measured the flow rate of liquid helium and deduced its superfluid velocity in a capillary flow experiment occurring in single $30~$nm long nanopores with radii ranging down from 20~nm to 3~nm. As the pore size is reduced towards the 1D limit, we observe: {\it i)}  a suppression of the pressure dependence of the superfluid velocity; {\it ii)} a temperature dependence of $v_{s}$ that surprisingly can be well-fitted by a powerlaw with a single exponent over a broad range of temperatures; and {\it iii)} decreasing critical velocities as a function of radius for channel sizes below $R \simeq 20$~nm, in stark contrast with what is observed in micron sized channels. We interpret these deviations from bulk behaviour as signaling the crossover to a quasi-1D state whereby the size of a critical topological defect is cut off by the channel radius.

\end{abstract}


\maketitle


Helium is the only known element in nature that becomes a superfluid, with its small mass and high symmetry cooperating to prevent solidification at atmospheric pressure as the temperature approaches absolute zero. For ${}^4$He, the ability to flow without viscosity below the $\lambda$-transition temperature, $T_\lambda$, is a paradigmatic manifestation of emergent phenomena and macroscopic quantum coherence, driven by both strong interactions and bosonic quantum statistics. Its superflow with velocity ${\vec v}_{s}=(\hbar /m){\vec  \nabla}\Phi$ is caused by a quantum-mechanical phase gradient of the wave function and {\it a priori} should only be limited by the Landau criterion of superfluidity, $v_\mathrm{L}$ due to the roton minimum in the helium excitation spectrum. However, years of experiments \cite{VAROQUAUX06} have shown that superfluid $^{4}$He exhibits a critical velocity that is well below $v_{\mathrm{L}}\simeq 60~\mathrm{m/s}$. The exact microscopic mechanism by which a superfluid dissipates energy remains a major unsolved problem in condensed matter physics. 


At a first glance, it would appear that this problem would only be exacerbated as the number of spatial dimensions decreases, as enhanced thermal and quantum fluctuations should push $T_\lambda\to 0$. However, in the one dimensional limit, the universal quantum hydrodynamics of Luttinger liquid theory \cite{HALDANE81,AFFLECK11} should apply, providing a host of theoretical predictions including the simultaneous algebraic spatial decay of both density-density and superfluid correlation functions. While there is a body of evidence of this exotic behaviour in low dimensional electronic systems \cite{ISHII2003,YACOBY2005,RITCHIE2009,LAROCHE2013} and ultracold low density gases \cite{CAZALILLA2011},  the analogous behaviour has yet to be confirmed experimentally in a highly-correlated bosonic fluid. Here, the physics of superflow should be qualitatively altered, with the superfluid density $\rho_s$ acquiring strong system size and frequency dependence \cite{EGGEL11}. Furthermore, neutral massflow transport properties should be strongly modified in one dimension, with the superfluid velocity $v_s$ exhibiting non-universal power law dependence on temperature and pressure. This crossover towards one dimension is manifest in the main findings of our work: (I) a suppression of the pressure dependence of $v_s$ for $R \simeq 3~\mathrm{nm}$ indicative of enhanced dissipation via phase slips, (II) a temperature dependence for $v_{s}$ that can be described by a powerlaw with a single exponent over a broad range of temperatures,  and (III) decreasing critical velocities as a function of radius for channel sizes below $R \simeq 20$~nm; behaviour strongly deviant from what is observed in micron sized channels.



  In this work, the mass flow rate of superfluid helium is measured in a capillary experiment through channels with radii as small as $R\simeq 3$~nm and lengths $L = 30~\mathrm{nm}$. To determine the effective dimensionality of this geometry, it is imperative to perform a comparative analysis of all possible relevant length scales. Unlike superconductors and superfluid $^{3}$He which undergo a BCS pairing, ${}^4$He has a very small coherence length, on the angstrom scale: $\xi_4(T)\simeq \xi_0(1-T/T_\lambda)^{-\nu}$, with $\xi_0 \simeq 3.45~\mathrm{\AA}$ and $\nu \approx 2/3$, making it technically difficult to fabricate a transverse confinement dimension with $R \ll \xi_4$ approaching the truly one dimensional limit, as, for example, $\xi_4\sim 0.5-1.5$~nm in the temperature range considered here. For $T = 0.5-2~\mathrm{K}$, $R$ can also be compared to the thermal de Broglie wavelength, $\Lambda(T) = \sqrt{2\pi \hbar^2/m k_\mathrm{B}T} \sim 1~\mathrm{nm}$ and a thermal length $L_T=\hbar c_1/k_B T \sim 1$~nm, where $c_1\simeq 235~\mathrm{m/s}$ is the first sound velocity of ${}^4$He. An alternative measure of one-dimensionality can be obtained by computing the thermal energy needed to populate transverse angular momentum states for a single helium atom confined inside a long hard cylinder of radius $R$: $T \sim \Delta_\perp/k_\mathrm{B} \simeq 3.5/R^2~\mathrm{nm^2}\cdot\mathrm{K} \sim 0.4~\mathrm{K}$ for $R = 3~\mathrm{nm}$. These estimates, which mostly neglect interaction effects, would place our flow experiments in a mesoscopic regime, with confinement length and energy scales on the order of the intrinsic ones in the problem. However, recent \emph{ab initio} simulations of ${}^4$He confined inside nanopores \cite{KULCHTSKYY,POLLET2014} have demonstrated that classical adsorption behaviour leads to an effective phase separation, between a quasi-1D superfluid core of reduced radius and concentric shells of quasi-solid helium near the pore walls. This effect, which is likely also present in our channels, would tend to provide additional confinement, allowing us to approach an effectively quasi-one-dimensional state.




Previous investigations of helium confined at the nanometer scale have focused on porous media such as in Vycor \cite{REPPY1999} and more recently in the zeolites and other mesoporous media. These studies have shown a possible new thermodynamic phase of $^{4}$He stabilized at low temperature \cite{TODA2007} as well a nuclear magnetic resonance signature (NMR) of a one-dimensional crossover for $^{3}$He \cite{YAGER2013}. While these advances are certainly considerable in the search for a strongly-interacting 1D neutral quantum liquid, our approach differs much in spirit from those cited above. In our experiment, the helium atoms are confined inside a {\it single, nearly cylindrical} pore, rather than in an extremely large number of them necessary to gain signal for a macroscopic probe. This lone pore, or channel, is tailor-made from a pristine Si$_3$N$_4$ membrane that can be fabricated with radii ranging from $R \sim1-100~\mathrm{nm}$.  The main advantage of our approach is that  there is no ensemble averaging over pore distributions and/or potential defects of the sample. Its main drawback, however, is that traditional bulk measurement techniques, such as specific heat or NMR most likely cannot be performed in a single nanopore containing only $\sim$10$^{4}$ to $10^{5}$ helium atoms. Taken as a whole, these two approaches are complementary to one another and similar in spirit to ``bottom-up {\it vs.} top-down'' or ``single-molecule {\it vs.} ensemble averaged'' studies in other fields, such as nanoelectronics or molecular biology.

%
\begin{figure}[t]
	\centering
		\includegraphics[width=1\linewidth]{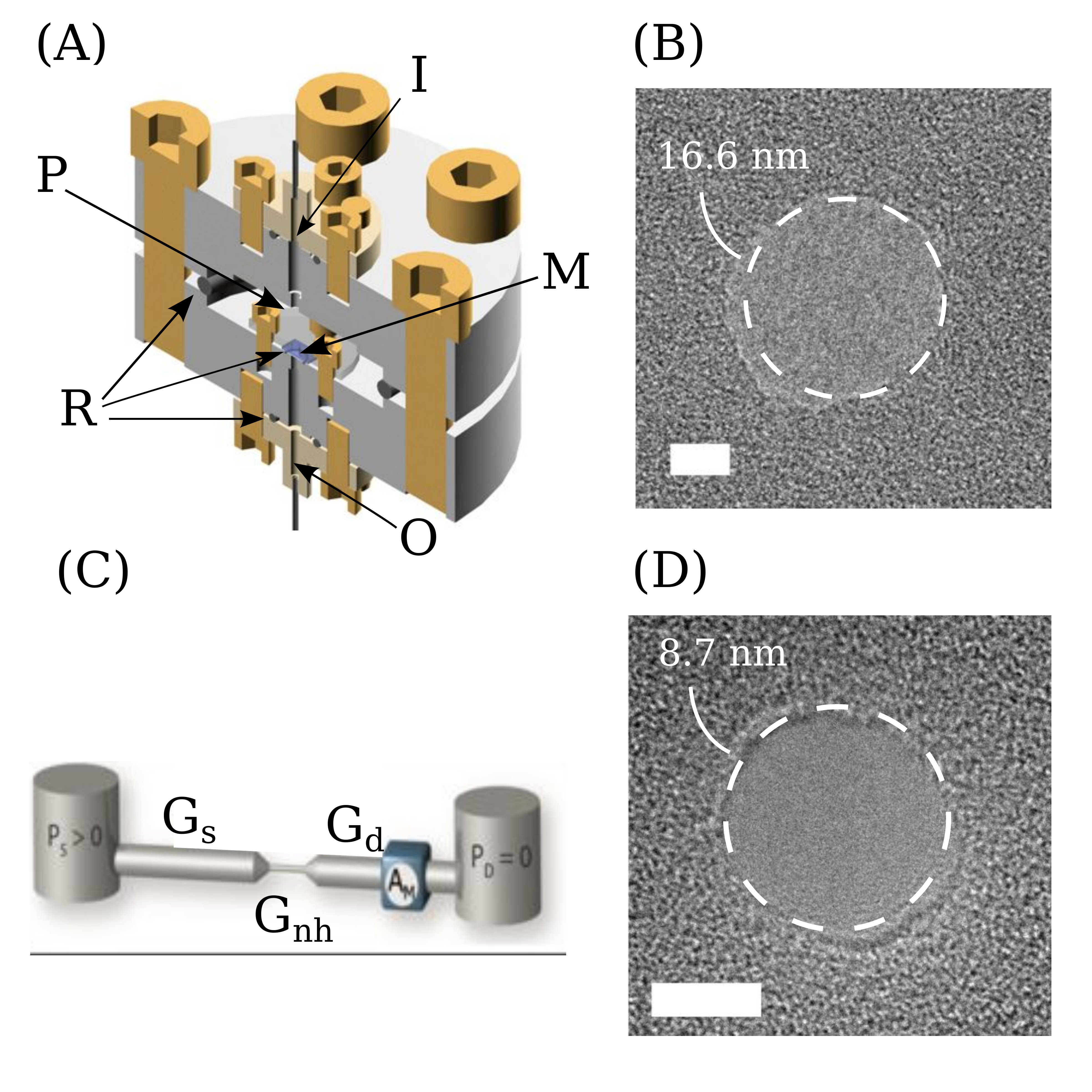}
	\caption{{\bf Design of the capillary flow experiment}. {\bf (A)} CAD drawing of the coin silver experimental cell. The inlet (I) and outlet (O) reservoir are connected to the top and bottom parts of the cell, and sealed with indium o-rings (R). The SiN membrane (M) is itself sealed to the bottom part of the cell with an indium o-ring and a push-on plate (P). {\bf (B)} Illustration of the flow experiment where the source reservoir is kept at a pressure $P_{S}>P_{D}\simeq 0$ and the flow measured by mass spectrometry ($A_m$) in a series experiment. {\bf (C)} and {\bf (D)} TEM picture of two nanopores used in this experiment. The bars represent 5 nm in both pictures. The diameter shown here is only an upper bound since the pore undergoes relaxation and its diameter decreases in size post-fabrication. In the experiment, the experimental pore radius was determined {\it in situ} using both Knudsen effusion in the gas phase and viscous flow measurements in the normal phase of liquid helium.  }
	\label{fig1}
\end{figure}

\begin{figure}[t]
	\centering
		\includegraphics[width=1.1\linewidth]{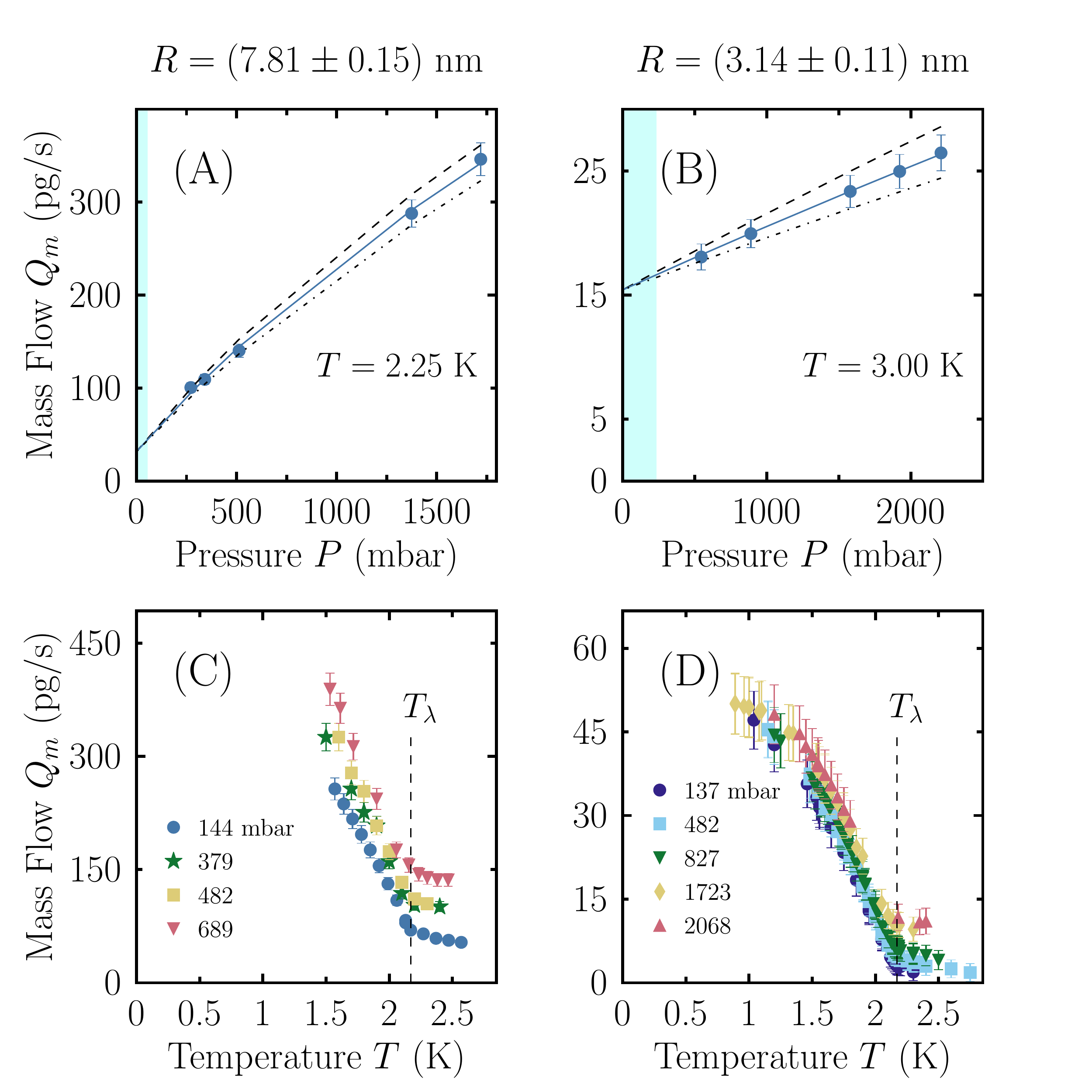}
	\caption{{\bf Flow measurements raw data.} {\bf (A)} Mass flow measurements as a function of pressure for a 7.81 nm  and {\bf (B)} a 3.15 nm pore radius in the normal state. The blue line is a fit of the data using Eq.\eqref{eq:normal_flow} and the dashed and dash-dotted lines are one standard variation from the mean value for the radius, with all other parameters kept constant. The finite intercept value at zero pressure is a spurious signal (see text). {\bf (C)} and {\bf (D)} Temperature dependence of the mass flow at several pressures. The dashed line shows the known superfluid transition temperature ($T_{\lambda}$) at saturated vapour pressure.}
	\label{fig2}
\end{figure}
%

The experiment is configured in a similar fashion, and follows the same procedure as previously reported in Ref. ~\cite{SAVARD2011}. However, the present work is performed in a newly-designed experimental cell made out of coin silver and shown in Fig.~\ref{fig1}(A).  The single nanopores were fabricated in the Si$_{3}$N$_4$ membrane using an electron beam from a Field-Effect Transmission Electron Microscope (FE-TEM), with images taken shortly after fabrication shown in Fig.~\ref{fig1}. While the single pores have a well-defined diameter, we have observed in previous work that their structure has a tendency to relax at room temperature, with the pore radius decreasing as a function of time (see supplementary information). To circumvent the uncertainty in the pore dimension, Knudsen effusion measurements in the gas phase of helium were conducted at low temperature (77~K) using the protocol discussed in Ref. ~\cite{SAVARD2009}.  The respective values obtained for each of the pores were determined to be $R^{\rm Kn}=8.2\pm0.5$ nm and $R^{\rm Kn}=3.10\pm 0.35$ nm (see supplementary information).

%

In a second step, the experimental cell was cooled down to liquid helium temperature (below 4.5~K).  Above $T_{\lambda}$, in the normal phase of helium,  the flow through the nanopore is viscously dissipative, and expected to follow the model developed for a short pipe by Langhaar \cite{LANGHAAR42}. In this phase, we have conducted pressure sweeps  at constant temperature while monitoring the mass flow rate $Q_m$,  as shown in Fig.~\ref{fig2}(A) and (B). In the absence of a chemical potential difference, the mass flow rate should go to zero. However, we observe a spurious signal as $\Delta P \to 0$ arising from evaporation at the walls of the channel. To determine this offset, the data were fitted with the flow equation for short pipe, 
\begin{equation}
\label{eq:normal_flow}
Q_n=\dfrac{8\pi\eta L}{\tilde{\alpha}}\left(\sqrt{1+\dfrac{\tilde{\alpha} \rho R^4}{32\eta^2 L^2}\Delta P}-1\right)
\end{equation}
where $\eta$ is the viscosity and $\tilde{\alpha}$ is a coefficient to take into account the acceleration of the fluid at the pipe end (see supplementary information).  In Fig.~\ref{fig2}~(A) and (B), the solid line is a fit to to the data with a radius of $R^{{\rm He_I}}=7.81\pm0.15$~nm and $R^{\rm He_I}=3.14\pm0.11$~nm.  These values are in excellent agreement with those determined independently via  Knudsen effusion measurements. Importantly,  it demonstrates {\it de facto} that our experiment can quantitatively determine the mass flow near the $\lambda$-transition in very small channels.



\begin{figure}[t]
	\centering
 \includegraphics[width=.88\linewidth]{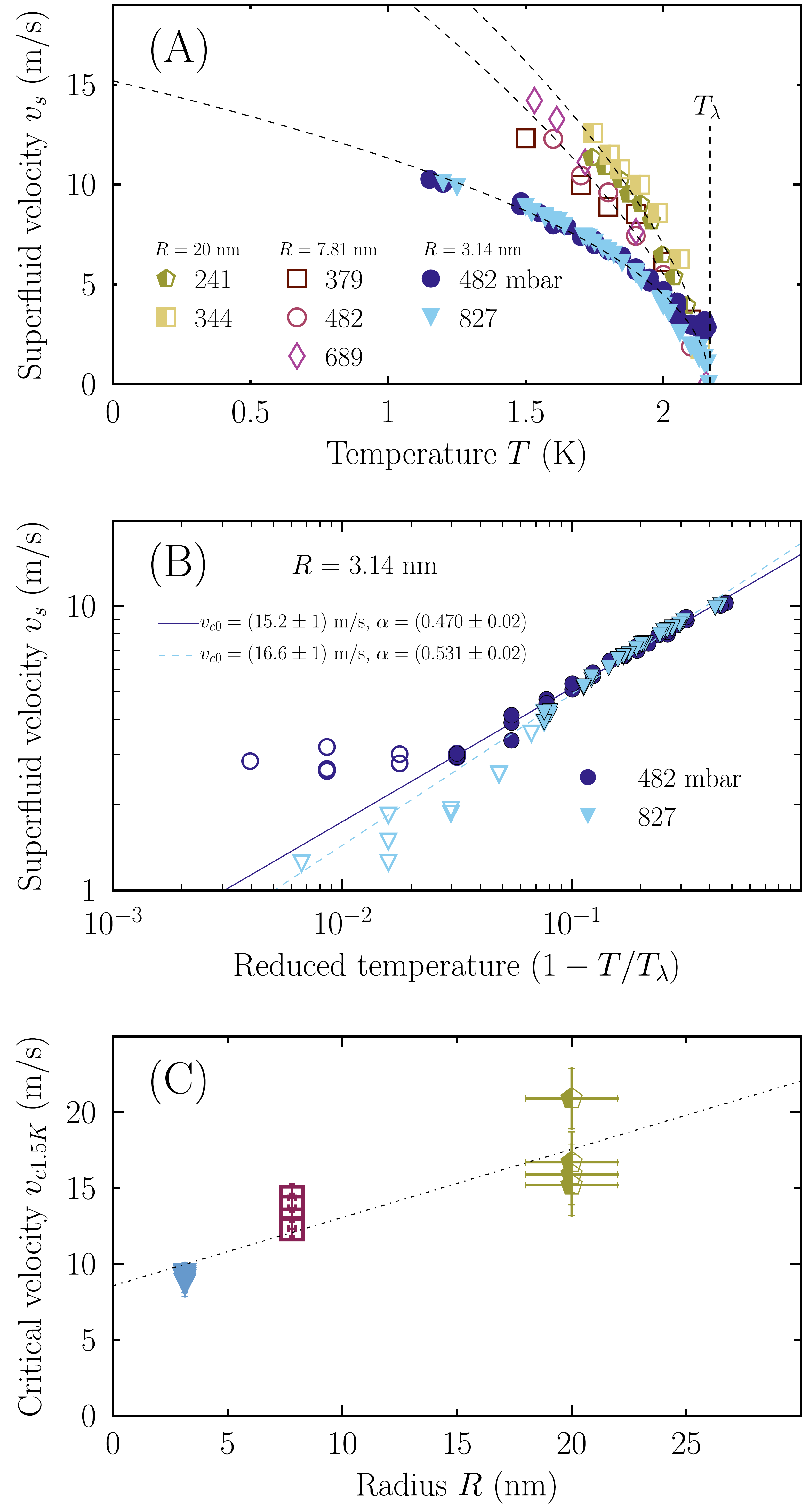}
	\caption{{\bf Superfluid velocities.} {\bf (A)} The superfluid velocities are shown at several pressures below 1 bar for the 7.81 nm pore (open symbols) and 3.14 nm (closed symbols), and 20 nm pore (half-filled symbols)\cite{SAVARD2011}. The dashed lines are fit using the powerlaw $v_{s}(T)=v_{c0}(1-\frac{T}{T_{\lambda}})^{\alpha}$ (see text). {\bf (B)}  Log-log plot of the superfluid velocity  versus the reduced temperature for the 3.14 nm pore data. The closed symbols indicate the data used for the power-law fit of the data taken at 482 (solid line) and  827 (dashed line) mbar pressure.   {\bf (C)} Critical velocity in  three nanopores extracted at 1.5K temperature (at several pressures less than 1 bar) in order to compare with previous work in much larger channels (see supplementary information). The superfluid velocities are assumed to be reaching the critical velocity. The dotted line is a blind linear fit shown here only as a guide-to-the-eye.}
	\label{fig3}
\end{figure}

The mass flow was measured as a function of temperature across the superfluid phase transition $T_{\lambda}$ at several pressures for both pores. These data are displayed in Fig.~\ref{fig2}~(C) and (D) with the offset previously discussed subtracted. Previous work in Vycor \cite{REPPY1999} have found the superfluid transition to be suppressed to 1.95K, however, 
the superfluid transition in our channels is observed at the temperature corresponding to the bulk value, 2.17K.  This is not surprising since we measure the total conductance of the nanopore channel {\it and} of the source reservoir in series, so the onset of superfluidity in the bulk is first observed at  $T_{\lambda}$. Considering only data below $T_\lambda$, we can extract the superfluid velocities using the two-fluid model where we assume $Q_\mathrm{tot}=Q_n+Q_s=(\rho_n v_n + \rho_s v_s)\pi R^2$. Subtracting $Q_{n}$ from the total mass flow using Eq.~\eqref{eq:normal_flow} yields the superfluid portion of the flow with a velocity $ v_s=Q_{s}/\pi R^2 \rho_s$. The superfluid density is taken from the bulk, as justified by previous work in Vycor (with a similar network pore size), albeit with a lower transition temperature \cite{REPPY1999}. The extracted superfluid velocities are shown in Fig.~\ref{fig3} for the lower pressure datasets, where linear response is expected to be a better approximation, and where the datasets were taken over a large range of temperatures. An inspection by eye readily shows that the superfluid velocities are smaller in the $R\simeq 3~\mathrm{nm}$ pore at similar pressures and temperatures. Such suppression of the flow velocity as the radius is decreased is in stark contrast with the bulk behaviour and shows that dissipation is increasing as the radius of the pore approaches a few nanometers. 
%

%

Near the bulk superfluid transition, it is well-established that the superfluid density follows a universal powerlaw form $\rho_s = \rho_0(1-T/T_\lambda)^{\nu}$, where $\nu$ is a correlation length critical exponent  found experimentally to be close to ${2}/{3}$.  Considering a slowly-varying quantum-mechanical wavefunction with a phase $\Phi$, the kinetic energy of the superfluid is given by $\rho_s v_s^2/2=\rho_s (\hbar^2/2m^2)|\vec{\nabla} \Phi|^2$. From scale invariance, we expect that near $T_\lambda$, the mean square of the superfluid velocity should scale with the correlation length $\xi_4(T)$ as $\overline{v_s^2}\sim 1/\xi_4(T) ^2\sim (1-{T}/{T_{\lambda}})^{2\nu}$. This result is strictly speaking valid only at temperatures very closed to $T_\lambda$, $(1-{T}/{T_{\lambda}})\lesssim 0.1$. From this hyperscaling anaysis, there is no reason to expect powerlaw behaviour in the superfluid velocity over a wide range in temperature away from $T_\lambda$. However, in the data shown in Fig.~\ref{fig3}(A), a powerlaw $v_s(T)=v_{c0}\left(1-{T}/{T_{\lambda}} \right)^{\alpha}$, where $v_{c0}$ is the superfluid critical velocity at $T=0~\mathrm{K}$, was used to fit all the data. A log-log plot of $v_s$ versus the reduced temperature is shown in Fig.~\ref{fig3}(B) for the 3.14 nm pore.  For this radius, where very little pressure dependence on the flow is observed, the powerlaw yields an exponent $0.53\pm 0.02$
and $0.47\pm 0.02$ for the low (482 mbar) and higher (827 mbar) dataset, respectively, and their critical velocity at zero temperature are $v_{c0}=15.2\pm1$~m/s and $v_{c0}=16.6\pm1$ m/s.  In contrast, the larger pore (7.81 nm) displays a significantly distinct exponent  $0.66\pm 0.05$ and zero-temperature critical velocity $v_{c0}=30.1\pm2.4$~m/s.  While not a proof, given the limited range in temperature explored,  this non-universal powerlaw behaviour as the dimensionality is reduced  is consistent with expectations from quantum hydrodynamics in 1D where increased fluctuations should prohibit long range order. 

Other important features of the  flow data  not previously observed are (I) the extremely weak pressure dependence below $T_\lambda$ for the smaller pore, and (II) an overall decrease in critical velocity as the channel size is reduced, in contrast to the behaviour $v_c \sim {\hbar \over m R} \ln ({R \over a_0})$, with $a_0$ the size of the vortex core, predicted by Feynman and found in larger channels (see supplementary information). The former is a hallmark of the macroscopic phase coherence that exists in a superfluid phase, in sharp contrast with the Euler prediction of a classical inviscid fluid,  $v_s = \sqrt{2 \Delta P / \rho_s}$. Using the Gibbs-Duhem relation to convert a pressure to chemical potential difference, energy conservation dictates that there must exist a dissipation mechanism in the channel with a rate $\Gamma$ such that $ h \Gamma = \frac{m \Delta P}{\rho_s} - \frac{1}{2} m v_s ^2$. From our data,  it is clear that the dissipation rate must be flow (pressure) dependent. The question of how energy is dissipated in superfluids has a long history, beginning with the proposal of Anderson \cite{ANDERSON} that, in analogy with the Josephson effect in superconductors, a steady state non-entropic flow may be achieved at a critical velocity $v_{c}$ via a mechanism that unwinds the phase of the order parameter in quanta of $2\pi$.  Such ``phase slips'', occurring at rate $\Gamma$,  corresponds to a process whereby the amplitude of the order parameter is instantaneously suppressed to zero at some point along the channel, and can be driven by either thermal or quantum fluctuations.  Momentum conservation dictates that such events can only occur in the presence of broken translational invariance along the pore \cite{KHLEBNIKOV04}. 

Microscopically, dissipation occurs through the creation of quantized vortex rings, the topological defects of superfluid hydrodynamics.  In our experiments, the size of critical vortex ring, $R_c$ plays a crucial role, and it is determined by the equilibrium condition between the relative frictional force between the normal and superfluid component and the hydrodynamic forces acting on the ring in the presence of flow.  Energetically, this manifests as a competition between a positive vortex energy that scales linearly with radius and a negative kinetic core energy scaling like its area. Langer and Fisher \cite{LANGER67} found $R_c \sim 3~\mathrm{nm}$ below $T_\lambda$, exactly the length scale of the smallest pore considered here.  When $R < R_c$, the maximum size of a vortex ring is constrained by the radius of the channel, and thus the energy barrier for their creation is lowered, leading to increased dissipation and an upper bound on $v_s$ set by the Feynman critical velocity. The suppression in the observed critical velocity at $T=1.5~\mathrm{K}$ as a function of decreasing radius shown in Fig.~\ref{fig3} can then be interpreted as a crossover to a regime where flow is dominated by the physics of the channel.  As the channel radius continues to decrease further, it is expected that backscattering of helium atoms at low temperature in the guise of quantum phase slips will increase, resulting in a continued suppression of the critical velocity.

This argument does not address the actual rate, or probability per unit space time that topological defects are created, and experimental estimates of $\Gamma$ were first made by Trela and Fairbank \cite{Trela}, who found $\Gamma \sim 1~\mathrm{Hz}$ for superfluid flow through constrictions with $R \sim 10^{-4}~\mathrm{m}$.  For the nanoscale pores considered here, we estimate that $\Gamma \sim 3-5~\mathrm{GHz}$,  well below the flow rate of $7.5\times10^{12}$~atoms/s measured in our smaller pore, yet approaching the quantum of mass flow $q=m^2/h \sim 10^{10}$~atom/s at one bar differential pressure and fluid density taken at saturated vapour pressure.



The behaviour of superfluid helium flow was studied in capillary channels down to $\sim$3 nm radius. For the smaller pore, the superfluid velocity can be well described by a powerlaw and it was found to be significantly smaller than in larger channels. This likely signals the crossing over to a quasi-one-dimensional state whereby increased fluctuations and interaction renormalization are modifying superfluidity.  As the channel size is reduced even further, near, or into the sub-nanometer range, we expect to observe physics characteristic of the truly one-dimensional limit.  In this regime, the algebraic decay of the superfluid order parameter will manifest itself as a reduction in the superfluid density as a function of channel length and the appearance of non-universal powerlaws in the massflow dependence on pressure $(Q_{1D}\sim \Delta P^\beta$) and temperature $(Q_{1D}\sim T^\gamma$) .  Such observations would be strikingly different than that seen due to the macroscopic quantum coherence of bulk helium, and would signal the experimental discovery of a one-dimensional bosonic quantum fluid.



We thank I. Affleck for illuminating discussions and comments. We also thank the (CM)$^2$ facility at Polytechnique Montreal for providing access and help with the TEM. This work was funded by  Natural Sciences and Engineering Research Council of Canada (NSERC), the Fonds Qu\'eb\'ecois de la Recherche sur la Nature et les Technologies FRQNT (Qu\'ebec) and  the Canadian Institute for Advanced Research (CIFAR). All data, analysis details and material recipes presented in this work are available upon request to G.G.



\newpage
\newpage
\section{SUPPLEMENTARY INFORMATION}

\subsection{Flow measurements}

\subsubsection{Design}

 Figure 1(A) shows a CAD drawing of the experimental cell used for the gas flow measurements and Fig. 1(C) and (D) shows a field-emission transmission electron microscope (FE-TEM) images of the nanoholes used. The $Si_3N_4$ wafer is installed in a coin silver cell and sealed by an indium o-ring  separating two reservoirs (inlet and outlet) in an experimental cell designed such that any mass transfer between the two reservoirs is restricted to occur through the nanohole. Capillaries connect the extremities of the experimental cell to a gas handling system such that pressurized helium can be introduced in the cell, flow through the nanohole, and be pumped from the outlet of the cell by a mass spectrometer, see Fig. \ref{fig1}(B). We use packed silver powder heat exchangers to condense helium before it enters the inlet of the cell and to ensure a good thermal anchor to the cryostat. The temperature was determined using two calibrated ruthenium oxide thermometers apposed on the experimental cell and the ${}^3$He pot. The temperature control was achieved with the PID loop of a LakeShore 340 AC resistance bridge. The coordination of the measurement and the PID control were handled in a homemade Python interface.  
 
 The drain pressure below the membrane ($P_{D}$) is kept at vacuum through continuous pumping and helium gas is introduced in the top part of the cell creating a pressure gradient $\Delta P = P_{S}-P_{D}\simeq P_{S}$ which  induces a mass flow $Q_m$. This flow was detected with a Pfeiffer vacuum Smart Test HLT560 calibrated with an external standard leak of $2.79\times10^{-8}$ atm$\cdotp$ cc/s $\pm 10-15$\%. A cartoon representation of the whole experiment is shown in Fig. 1(B). The two reservoirs are depicted by capillary conductances $G_{S}$ and $G_{D}$ in series before and after the nanohole with a conductance $G_{nh}$.  The mass spectrometer is denoted by $A_{M}$ and measures the volumetric flow when the drain side of the set-up is kept under  vacuum, typically below $\sim 2\cdotp{10^{-3}}$ mbar. With our technique, the total conductance $G_{T}^{-1}=G_{S}^{-1}+G_{D}^{-1}+G_{nh}^{-1}$ of the circuit  is measured. The source and drain conductance can be estimated using the infinite pipe approximation for Poiseuille flow ($G_S\sim  10^{-11}$ m$\cdot$s at $\sim 1$ bar) and Knudsen free-molecular diffusion ($G_{D}\sim  10^{-13}$ m$\cdot$s  at $10^{-3}$ mbar). These conductances are several orders of magnitude larger than the nanohole conductance which has a typical value $G_{nh} \sim 10^{-18}$ m$\cdot$s (see Fig. 2(B)).  We can therefore neglect the source and drain conductance to a very good approximation.

\begin{figure}[t]
	\centering
		\includegraphics[width=.9\linewidth]{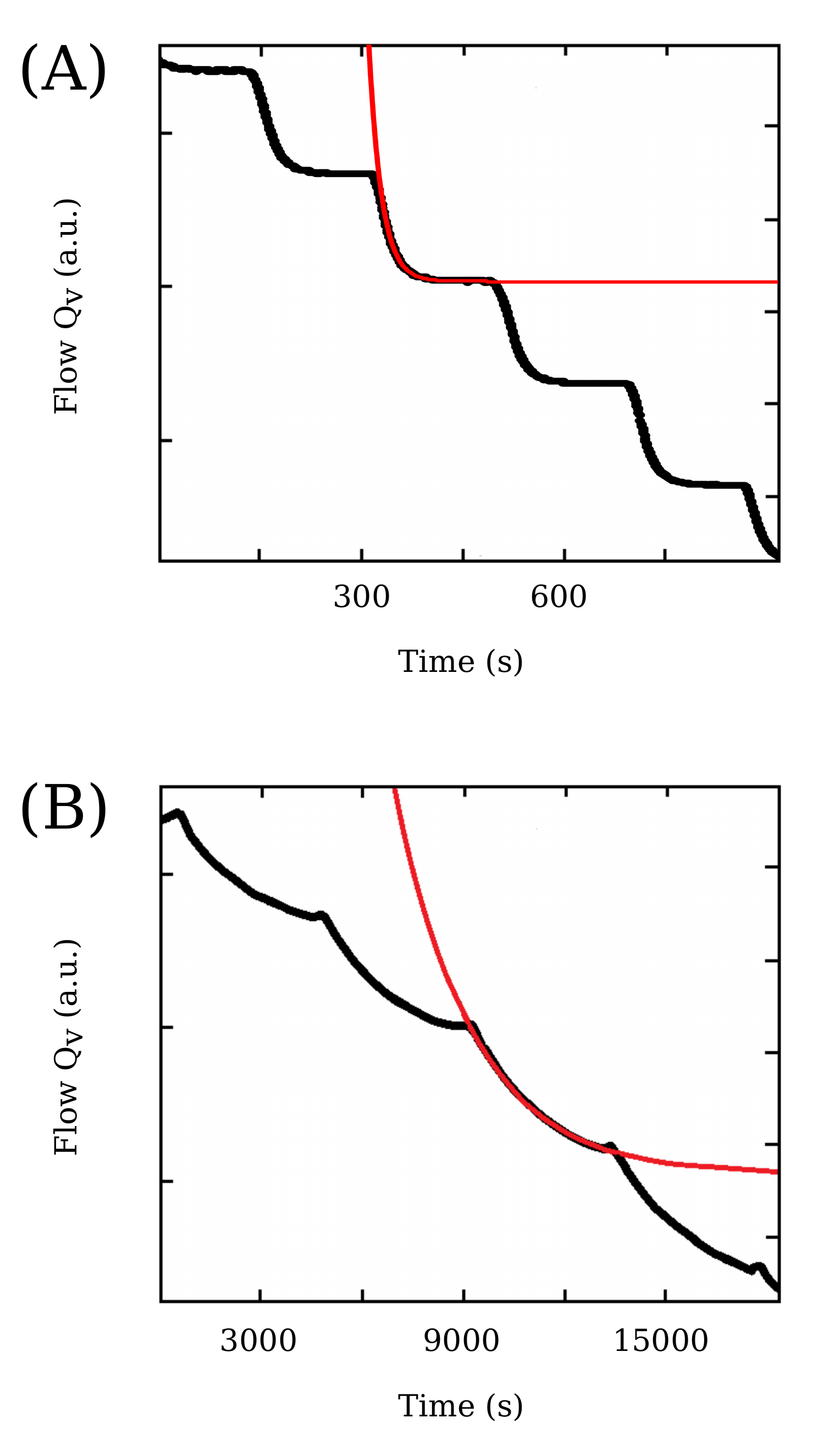}
	\caption{{\bf Flow measurements}.  {\bf (A)} Volumetric flow through a single 101 nm nanopore at 77K in the gas phase of helium when the pressure differential is decreased in a stepwise fashion. The red line is a fit of the function $Q_f(t,P_f)=Q_i(P_i)+\Delta Q e^{-t/\tau}$ used to extract the equilibrium value. {\bf (B)} Similar measurements in the 6 nm diameter nanopore for the superfluid phase of helium when the temperature is increased in a stepwise fashion. The red line is obtained in the same fashion as in (A).}
	\label{fig1supp}
\end{figure}

 \subsubsection{Experimental procedure}
 
	The procedure for making the measurements is as follows: we first empty both sides of the cell at a temperature well above the helium boiling point so as to ensure that no residual helium is present in either reservoir. The mass spectrometer is then connected to the outlet of the cell to determine a background signal that is treated as an offset to the pressure-driven flow of interest in this study. This background signal was found to be always less than $ \sim 5 \cdot 10^{-1}$ pg/s in the liquid phase and less than $ \sim 3 \cdot 10^{-2}$ pg/s in the gas phase. In the liquid phase it is less than the measured mass flow by a few orders of magnitude whereas in the gas phase at extremely low pressures it eventually becomes comparable to the flow signal. In  the next step, the whole apparatus is cooled below the $\lambda$-transition so that gaseous helium introduced from the gas handling system condenses and fills the heat exchanger and inlet of the experimental cell. Once condensation is achieved, the higher pressure above the membrane forces the liquid helium to flow through the nanohole.  When atoms reach the very low pressures in the drain reservoir, they evaporate and are pumped out to the mass spectrometer.  The volumetric flow signal is then monitored as the temperature of the cell is slowly increased. The measurement is then repeated at different pressure gradients across the nanopore. The volumetric flow is converted into a mass flow using $Q_m=(Q_v-Q_{v,bkg})/(10R_s T_{room})$. The factor of ten here comes from the transformation of liters to cubic meters and mbars to Pascals. Finally, $T_{room}$ is in Kelvin and the specific gas constant for helium $R_s=R/M_{m,He}$ is in Joules per kilogram Kelvin $J/(kg K)$. 

\subsubsection{Time constants and flow}

In analogy with an electrical circuit with a time constant $\tau=RC$,  the time required for the mass flow signal to stabilize upon a pressure of temperature variation is inversely proportional to the conductance of the nanohole. Figure \ref{fig1supp} shows measurements of the volumetric flow versus time and a fit of the signal with an exponential decay function of the form $Q_f(t,P_f)=Q_i(P_i)+\Delta Q e^{-t/\tau}$. The time constant $\tau$ for the superfluid flow through the smallest nanopore  is typically of order of 2000 seconds. We have verified that waiting over a period of time longer than $2\tau$ did not improved the accuracy of the fit in a significant fashion.

\begin{figure}[t]
	\centering
		\includegraphics[width=1.\linewidth]{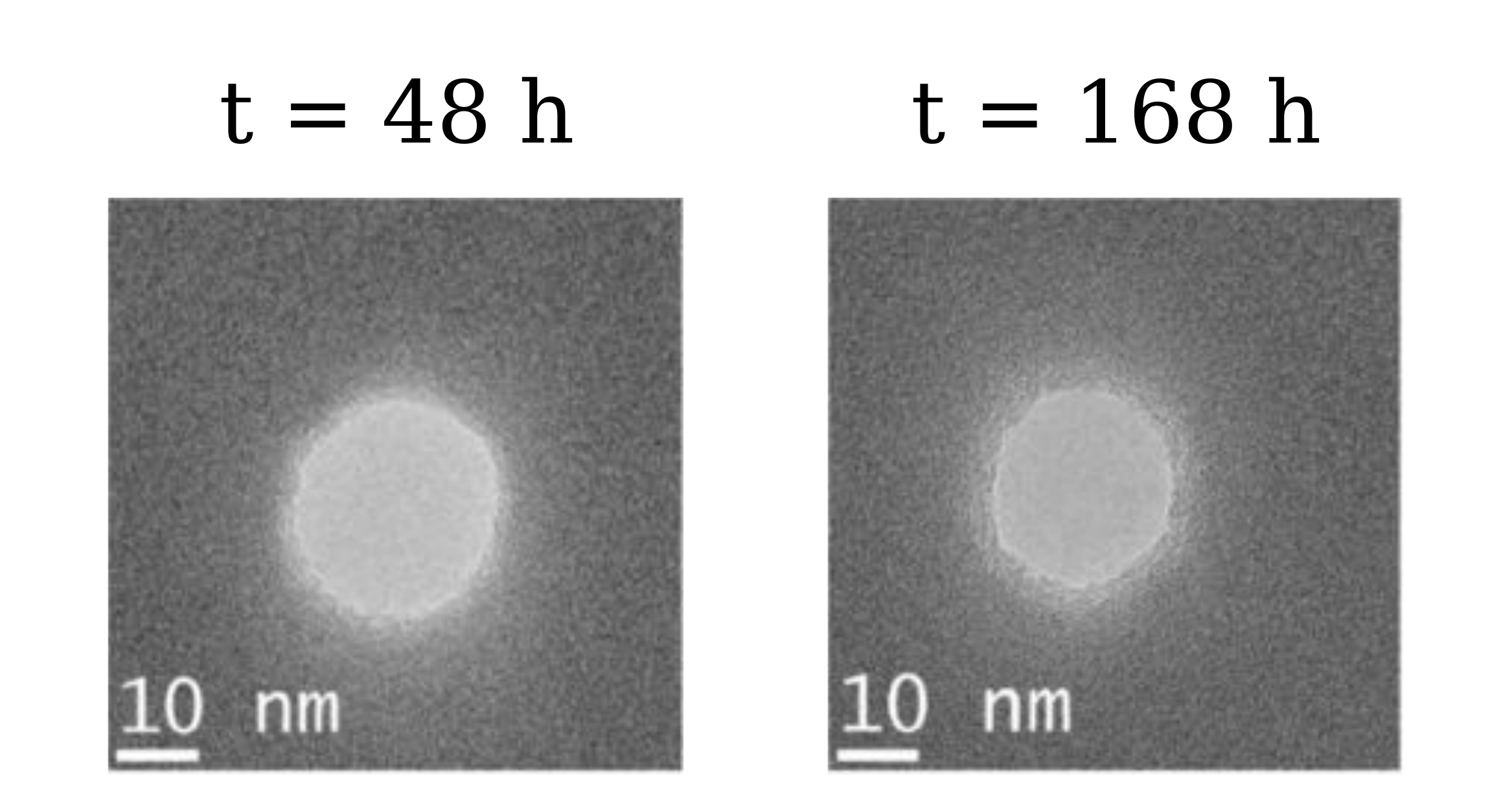}
	\caption{{\bf Nanopore structural stability}. The TEM image shows the nanopore at several days apart while kept at room temperature in a clean environment. This structural relaxation has been observed in several samples with  different $R/L$ ratios.  }
	\label{fig2supp}
\end{figure}

\subsection{Radii determination}

\subsubsection{Nanopore structural stability}

The nanopores used in the present study were fabricated by focusing a TEM-FE beam on $30$ nm thick $Si_3N_4$ membranes as in Ref. \cite{SAVARD2009,SAVARD2011}. For the smaller nanopores, we have have found that it had a tendency to relax during post-fabrication. An example of such relaxation is  shown in Fig. \ref{fig2supp} where the nanopore was imaged at different times following the fabrication. While this structural relaxation is more acute for the smaller pore, we have found that the relaxation process stopped at cryogenic temperature, below $\sim$5~K. This was verified by performing Knudsen effusion measurements before and after long period of time during which the membrane was kept at helium temperature. However, because of the deadtime between the fabrication and the cooling procedure in the cryostat, this relaxation process causes an uncertainty in the radii determination of the pore. For this reason we have developed two independent ways to determine the radii {\it in situ} using both Knudsen effusion and classical fluid dynamics.

\subsubsection{Radius determination from Knudsen effusion}

\begin{figure}[t]
	\centering
		\includegraphics[width=.8\linewidth]{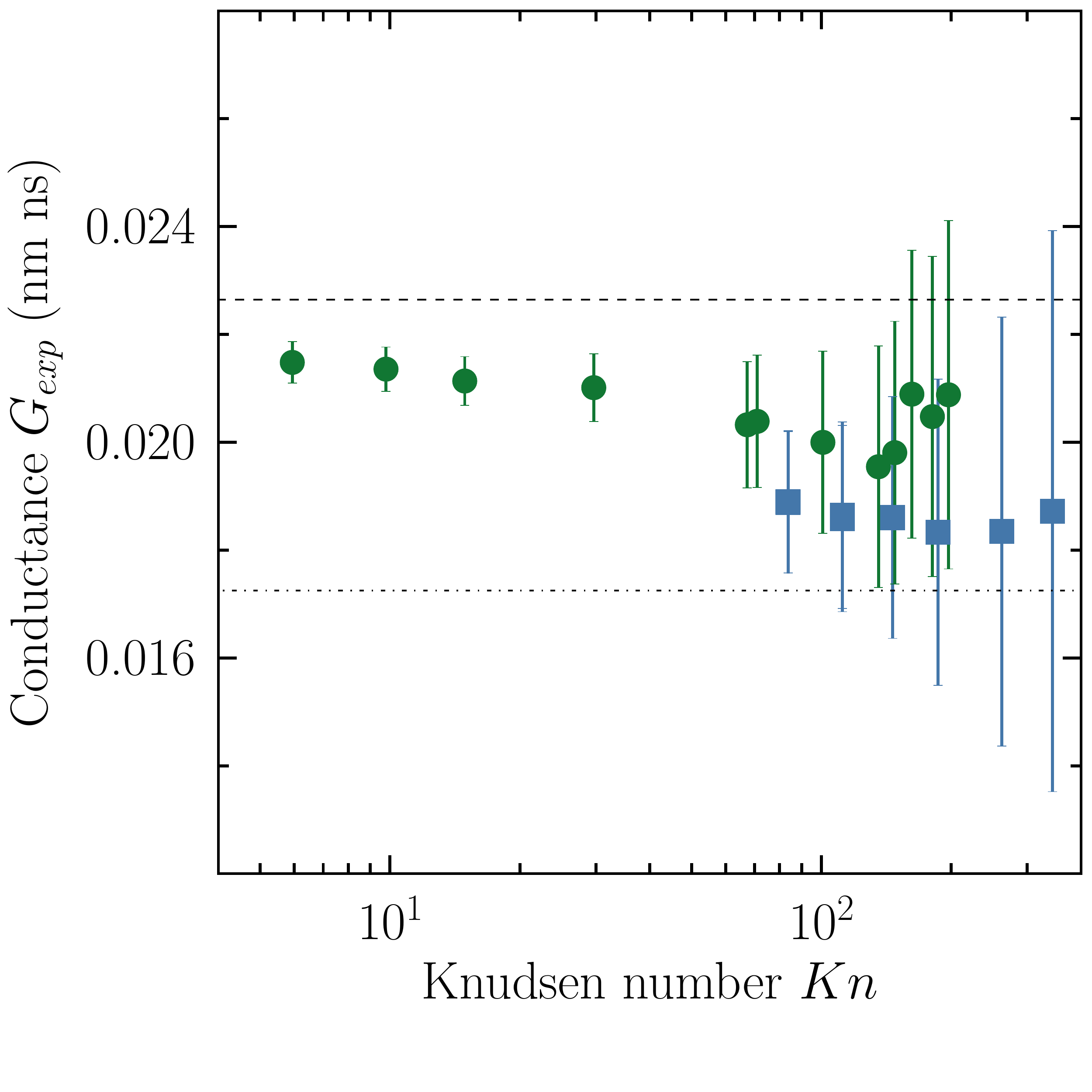}
	\caption{{\bf Determination of the radius by Knudsen effusion}. Conductance values obtained from the ratio of the measured mass flow and the applied pressure gradient for the smaller nanopore. The green circles and the blue squares are Knudsen effusion measurements at 77K made prior and after superfluid flow measurements, respectively, two months apart from each other. The dashed  and  dashed-dotted lines correspond to the maximum value of $G_{exp}+\delta G_{exp}$, and the minimum of $G_{exp}-\delta G_{exp}$, respectively. The data points are displayed here with three standard deviations.}.
	\label{SupInf_Knudsen}
\end{figure}

The methodology is similar to that reported in Ref. \cite{SAVARD2009} where the conductance of the nanopore is measured as a function of the Knudsen number (defined here as the ratio of the atom's mean free path to the nanopore diameter). For the smaller nanopore, the Knudsen number is sufficiently high that we can therefore neglect the contribution from the viscous regime. The Knudsen conductance is given by $G_{th}=R^2\kappa(R,L,\theta)\sqrt{\frac{\pi}{2R_sT_{cell}}}$
where $R$ is the radius of the nanopore, $L$ its length and $\theta$ the opening angle of the nanopore (when $\theta=0^{\circ}$ the nanopore is a cylinder). The opening angle of other nanopores with similar dimensions were measured using a TEM tomography technique in \cite{KIM2006} and was found to be close to 30$^{\circ}$. In our case,   an angle near  15$^{\circ}$ was found to best fit the Knudsen effusion data. While this angle is consistent with the profile reconstructed from the TEM picture,  the uncertainty in its precise determination  will lead to an uncertainty in the radius. The Clausing factor $\kappa(R,L,\theta)$  is a number between 0 and 1 that express the probability for an atom to go from one side of the nanopore to the other by bouncing on the walls. Figure \ref{SupInf_Knudsen} shows the value of the conductance as a fonction of Knudsen number $Kn$. The data at higher $Kn$ have larger uncertainties because they correspond to very low pressures/flow regimes. The radius of the nanopore and its uncertainty were extracted from the minimization of $G_{th}(R)-G_{exp}$ and $G_{th}(R)-(G_{exp}\pm \delta G_{exp})$  for which $L=30$nm and $\theta =(15 \pm 5 ^{\circ})$ and where $G_{exp}$ is the weighted average of the measured values. The deviation  $\delta G_{exp}$ here corresponds here to three standard deviation from $G_{exp}$. The extracted values for the radii are $R^{Kn}=8.2\pm 0.5$ nm and $R^{Kn}=3.1\pm 0.35$ nm.


\subsubsection{Radius determination from the viscous normal flow }

Pressure sweeps were performed in the normal phase of the liquid helium and the data were fitted against a slightly modified model of short pipe viscous flow from Langhaar \cite{LANGHAAR42},

\begin{figure}[t]
	\centering
		\includegraphics[width=.85\linewidth]{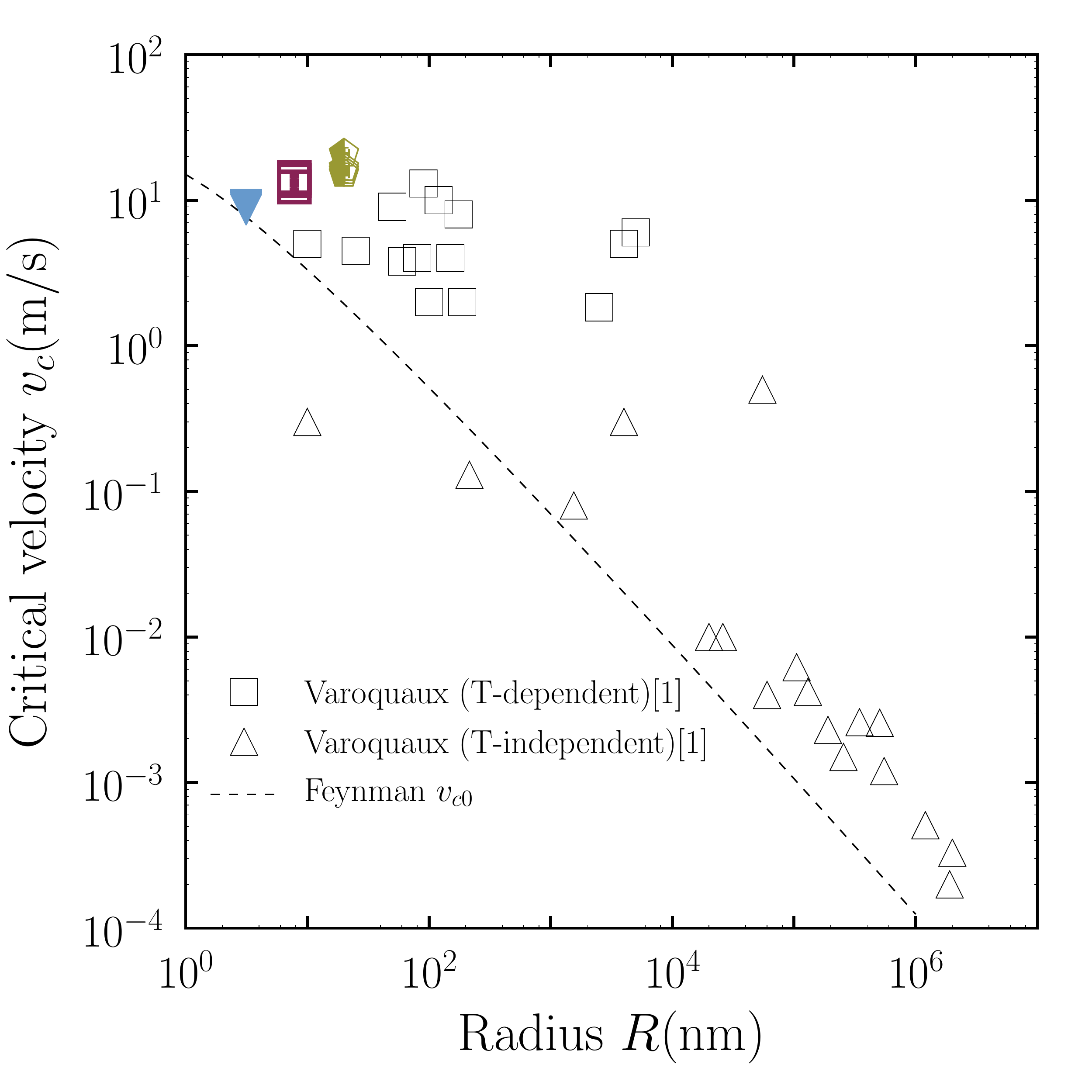}
	\caption{{\bf Critical velocities versus channel size}. Critical velocities at temperature $T=1.5$K from this work (colour) are displayed alongside with critical velocities from previous work (black and white) and summarized by Varoquaux \cite{VAROQUAUX06}.  The open squares correspond mainly to pressure driven AC flow experiment with circular nanopores in $\sim$$100$nm thick nickel foil and in nanoporous 5$\mu$m thick mica. The open triangles represent an heterogenous set of data collected from various type of experiments (heat flow, oscillations). The outlier triangle data point at small radius is a thin film experiment. Because these measurements were performed at temperature $T\simeq 0.95$K and above, the superfluid critical values is quoted at 1.5K.}
	\label{SupInf_critical}
\end{figure}

\begin{equation}
\label{eq:normal_flow_mod}
Q_m=\dfrac{8\pi\eta L}{\tilde{\alpha}}\left(\sqrt{1+\dfrac{\tilde{\alpha} \rho R^4}{32\eta^2 L^2}\Delta P}-1\right)+Q_{m,offset}
\end{equation}. 

The last term, $Q_{m,offset}$, is required here because we observe a spurious signal as $\Delta P \to 0$. This signal  is believed to arise from evaporation at the walls on the drain side. The free parameters in Eq.\eqref{eq:normal_flow_mod} are the radius $R$, the mass flow offset $Q_{m,offset}$, and $\tilde{\alpha}$ which is a geometry-dependent factor accounting for the acceleration of the fluid at the nanopore boundary. The best fit values were determined using a least squares method, evaluating $\sum(Q_{m,model}-Q_{m,meas})^2$ over a cube in parameter space in order to find a global minimum.  In Fig.~\ref{fig2}~(A) and (B), the solid line is a fit to to the data with a radius of $R^{He_I}=7.81\pm0.15$nm and $R^{He_1}=3.14\pm0.11$nm.  These values are in excellent agreement with those determined independently via the Knudsen effusion measurements discussed above. 

It is interesting to note that, as $\rho$ and $\eta$ are nearly constant in the normal phase, for sufficiently small values of $R$  the influence of the $\tilde{\alpha}$ parameter becomes negligible: $\frac{8\pi\eta L}{\tilde{\alpha}}\left(\sqrt{1+\frac{\tilde{\alpha} \rho R^4}{32\eta^2 L^2}\Delta P}-1\right) \simeq \frac{8\pi \eta L}{\tilde{\alpha}}\left(1+\frac{\tilde{\alpha} \rho R^4}{64\eta^2 L^2}\Delta P-1\right)=\frac{\pi \rho R^4}{8\eta L}\Delta P$. This is the case for the smaller nanopore of radius $R \simeq 3$ nm, but not for the larger nanopore of $R\simeq 8$ nm.  The parameter $\tilde{\alpha}$ was determined from a fit using Knudsen effusion data and  found to be equal to 4.7.  We have verified that modifications in $Q_{m,model}$ arising from a variation of $\tilde{\alpha}$  between 0 and 10 was negligable, {\it i.e.} the value of $\tilde{\alpha}$ had little or no influence on the data.


\begin{thebibliography}{25}

\bibitem{VAROQUAUX06}
E. Varoquaux, {\it C. R. Phys.} \textbf{7}, 1101 (2006).

\bibitem{HALDANE81}
 F. D. M. Haldane, {\it J Phys C: Solid State Phys} 14, 2585 (1981); {\it Phys. Rev. Lett.} {\bf 47}, 1840 (1981).
 
 \bibitem{AFFLECK11} A. Del Maestro, M. Boninsegni, and I. Affleck, {\it Phys. Rev. Lett.} \textbf{106}, 105303 (2011), and references therein.
 


\bibitem{ISHII2003} H. Ishii, H. Kataura, H. Shiozawa, H. Yoshioka, H. Otsubo, Y. Takayama, T. Miyahara, S. Suzuki, Y. Achiba, M. Nakatake, T. Narimura, M. Higashiguchi, K. Shimada, H. Namatame M. Taniguchi, {\it  Nature} {\bf 426}, 540 (2003).
 
\bibitem{YACOBY2005} O. M. Auslaender, H. Steinberg, A. Yacoby, Y. Tserkovnyak, B. I. Halperin, K. W. Baldwin, L. N. Pfeiffer, K. W. West, {\it Science} {\bf 308}, 88-92 (2005).

\bibitem{RITCHIE2009} Y. Jompol, C. J. B. Ford, J. P. Griffiths, I. Farrer, G. A. C. Jones, D. Anderson, D. A. Ritchie, T. W. Silk, A. J. Schofield, {\it Science} {\bf 325}, 597-601 (2009).

\bibitem{LAROCHE2013}D. Laroche, G. Gervais, M. P. Lilly and J. L. Reno,  {\it Science} {\bf 343}, 631-634 (2014). 


\bibitem{CAZALILLA2011}M. A. Cazalilla, R. Citro, T. Giamarchi, E. Orignac, and M. Rigol, {\it Rev. Mod. Phys}. {\bf 83}, 1405 (2011).


\bibitem{EGGEL11}T. Eggel, M. A. Cazalilla, and M. Oshikawa, {\it Phys. Rev. Lett.} \textbf{107}, 275302 (2011).

\bibitem{KULCHTSKYY} B. Kulchytskyy, G. Gervais, and A. Del Maestro, {\it Phys. Rev. B} \textbf{88}, 064512 (2013).


\bibitem{POLLET2014}
    L. Pollet and A. B. Kuklov, {\it Phys. Rev. Lett.} \textbf{113}, 045301 (2014).

%
 
 

\bibitem{REPPY1999} G. M. Zassenhaus and J. D. Reppy, {\it Phys. Rev. Lett.} {\bf 83}, 4800 (1999), and references therein.

\bibitem{TODA2007}R. Toda, M. Hieda, T. Matsushita, N. Wada, J. Taniguchi, H. Ikegami, S. Inagaki, and Y. Fukushima, {\it Phys. Rev. Lett.} \textbf{99}, 255301 (2007).




 \bibitem{YAGER2013} B. Yager, J. Ny\'eki, A. Casey, B. P. Cowan, C. P. Lusher, and J. Saunders {\it Phys. Rev. Lett.} {\bf 111}, 215303 (2013).
 
 
 
 
 
 \bibitem{SAVARD2011} M. Savard, G. Dauphinais and G. Gervais, {\it Phys. Rev. Lett.} {\bf 107}, 254501 (2011).
 
 
 
  \bibitem{SAVARD2009} M. Savard, C. Tremblay-Darveau and G. Gervais, {\it Phys. Rev. Lett.} \textbf{103}, 104502 (2009).
 
 
 
\bibitem{LANGHAAR42}
H. L. Langhaar, {\it J. Appl. Mech.} \textbf{9}, A55-A58 (1942).

 
\bibitem{ANDERSON}
P. W. Anderson, {\it Rev. Mod. Phys.} \textbf{38}, 298 (1966).

\bibitem{KHLEBNIKOV04}
    S. Khlebnikov, {\it Phys. Rev. Lett.} \textbf{93}, 090403 (2004).

\bibitem{LANGER67}
    J. S. Langer and M. E. Fisher, {\it Phys. Rev. Lett.} \textbf{19}, 560 (1967).

\bibitem{Trela}
    W. J. Trela and W. M. Fairbank, {\it Phys. Rev. Lett.} \textbf{19}, 822 (1967). 

 
 
 
 
 
 
 








































\end{thebibliography}

\begin{thebibliography}{25}

\bibitem{VAROQUAUX06}
E. Varoquaux, {\it C. R. Phys.} \textbf{7}, 1101 (2006).


\bibitem{KIM2006}
M. J. Kim, M. Wanunu, D. C. Bell, and A. Meller, {\it Advanced Materials} \textbf{18}, 3149-3153 (2006).

\end{thebibliography}

\subsection{Critical velocities}

In the two-fluid model proposed by Landau and Tisza the total mass current is given by $J_{total}=\rho_s v_s+\rho_n v_n$  and  total density is given by the sum of the superfluid and normal component $\rho=\rho_s+\rho_n$. Per symmetry of the pore, we consider the flow to be in the axial direction of the nanohole so  $Q_{m,total}=J_{total}\pi R^2$. The normal part is well-modeled by the function given in equation \eqref{eq:normal_flow}, with $\rho$ here replaced by $\rho_n$. The superfluid velocities are given by $v_s=Q_{s}/\pi R^2 \rho_s=(Q_{m,total}-Q_n)/\pi R^2. \rho_s$. The values from our work are reported in Fig. \ref{SupInf_critical}  at 1.5~K, per previous  convention in the literature. Finally, we have made the assumption that the superfluid velocities were reaching the critical velocities, {\it i.e.} the superfluid velocity was only limited by dissipation.


%

\end{document}